

\magnification=1200
\vbadness=10000
\hfuzz=10pt \overfullrule=0pt
\baselineskip=12pt
\parindent 20pt \parskip 6pt

\hfill {~}
\vskip .6in

\centerline{{\bf CONSEQUENCES OF PROPAGATING TORSION}}
\vskip .2cm
\centerline{{\bf IN CONNECTION-DYNAMIC THEORIES OF GRAVITY}
\footnote{$^*$}{This work
was supported in part by NASA under Grants no. NAGW-931 and
NGT-50850, by the National Science Foundation under grant
PHY/92-06867, and by the U.S. Department
of Energy (D.O.E.) under contract no. DE-FC02-94ER40818.}}
\vskip .3in

\centerline {Sean M. Carroll$^{(1)}$ and George B. Field$^{(2)}$}
\vskip .3cm
\centerline{\it $^{(1)}$Center for Theoretical Physics, Laboratory for
Nuclear Science}
\centerline{\it and Department of Physics}
\centerline{\it Massachusetts Institute of Technology}
\centerline{\it Cambridge, Massachusetts\quad 02139}
\centerline{{\it email: carroll@marie.mit.edu}}
\vskip .3cm
\centerline{\it $^{(2)}$Harvard-Smithsonian Center for Astrophysics}
\centerline{\it Cambridge, Massachusetts\quad 02138}
\centerline{{\it email: field@cfa.harvard.edu}}

\vskip .3in

\centerline{\bf Abstract}
\vskip .1in

We discuss the possibility of constraining
theories of gravity in which the connection is a fundamental variable
by searching for observational consequences
of the torsion degrees of freedom.  In a wide
class of models, the only modes of the torsion tensor which
interact with matter are either a massive scalar or a massive
spin-1 boson.  Focusing on the scalar version,
we study constraints on the two-dimensional parameter space
characterizing the theory.  For reasonable
choices of these parameters the torsion decays quickly into
matter fields, and no long-range fields are generated which
could be discovered by ground-based or astrophysical experiments.

\vfill

\centerline{CTP~\# 2291 \hfill March 1994}

\centerline{gr-qc/9403058 \hfill}
\vskip .1in

\eject
\baselineskip=14pt

\noindent{\bf I. Introduction}

General relativity as formulated by Einstein describes the dynamics of
a metric tensor field $g_{\mu\nu}$ and the response of matter to this
metric.  Covariant derivatives are taken with respect to the
Christoffel connection on the tangent bundle.  This is the unique
metric-compatible and torsion-free
connection, and is treated as a quantity derived from the metric.
Despite the success of Einstein's theory in passing observational
tests, there are strong indications that general relativity is
incomplete --- the prediction of singularities, and especially the
difficulty of formulating a quantum theory.  It is therefore natural
to explore modifications of general relativity in which these
problems may be overcome.  One popular modification is to make
the connection itself a fundamental variable in its own right,
rather than a convenient expression for a certain function of the
metric.

Such ``connection-dynamic'' theories can take different forms.
A simple approach is the first-order or Palatini formulation
of conventional gravity [1].  The action of this theory
is that of general relativity, with the connection varied independently
rather than given as a function of the metric.  The
resulting equations of motion lead to the usual expression for the
connection in terms of the metric, with extra terms depending
algebraically on the matter fields.  Extensions of this procedure
have been adopted in attempts to construct quantum versions of
general relativity, including work in (3+1) dimensions [2]
and (2+1) dimensions [3].  (For reviews
and other approaches see [4].)  The additional terms contributing to
the connection in these theories
are characterized by the torsion tensor; since no derivatives of
the torsion appear in the Ricci scalar, the Palatini formulation
leads to non-propagating torsion [5].

Once torsion has been introduced by taking the connection to be an
independent variable, the restriction to non-propagating torsion
arises as much from historical accident as from first principles.
Just as the Einstein-Hilbert action provides dynamics for the
metric degrees of freedom, it is straightforward to consider the
addition of extra terms to the action which would provide
dynamics for the torsion degrees of freedom.

A comparison with conventional gauge theories serves to illustrate
why we believe that
a dynamical torsion tensor is a natural expectation when
the connection is treated as a variable independent from the metric.
In gauge theories of an internal symmetry, the connection is
specified by a non-gauge-invariant vector potential with an
associated gauge-covariant tensor, the curvature or field
strength.  In contrast, the connection $\nabla$ on spacetime is
associated with two tensors, the curvature and the
torsion.\footnote{$^1$}{In general, the connection $\nabla$ on the
tangent bundle is a GL(4,{\bf ~R}) connection.
When we restrict our attention to metric-compatible connections the
group is reduced to SO(3,1).}
We may specify the curvature and torsion of $\nabla$ in
terms of their action on vector fields $X$, $Y$, and $Z$.  (For
additional formulae see [5,6].) The curvature, a $(1,3)$ tensor, is
  $$R(X,Y)Z\equiv\nabla_X\nabla_YZ-\nabla_Y\nabla_XZ-\nabla_{[X,Y]}Z,
  \eqno(1.1)$$
while the torsion, a $(1,2)$ tensor, is
  $$T(X,Y)\equiv\nabla_XY-\nabla_YX-[X,Y].\eqno(1.2)$$
In these expressions we use $\nabla_X$ to denote a covariant
derivative in the direction along $X$, and $[X,Y]$ for the Lie
bracket.\footnote{$^2$}{As can be seen from (1.2), in the definition of
  torsion (unlike that of the curvature) a single vector field such
  as $X$ or $Y$ serves both as a direction in spacetime along which a
  covariant derivative is taken and as the object being differentiated.
  This is clearly only possible when the vector field is a section
  of the tangent bundle rather than an ``internal'' vector bundle;
  thus, the existence of torsion distinguishes the connection on the
  tangent bundle from the connections
  familiar from conventional gauge theories.}
In a basis adapted to a set of coordinates $x^\mu$, we can decompose
these tensors in terms of their components; thus the covariant
derivative is
  $$\eqalign{(\nabla_XY)^\lambda&=X^\mu\nabla_\mu Y^\lambda\cr
  &=X^\mu(\partial_\mu Y^\lambda+\Gamma^\lambda_{\mu\nu}Y^\nu),\cr}
  \eqno(1.3)$$
while the Lie bracket is
  $$[X,Y]^\mu\equiv X^\nu\partial_\nu Y^\mu-Y^\nu\partial_\nu X^\mu.
  \eqno(1.4)$$
The curvature tensor is then
  $$R^\alpha{}_{\beta\mu\nu}=\partial_\mu\Gamma^\alpha_{\nu\beta}
  -\partial_\nu\Gamma^\alpha_{\mu\beta}+\Gamma^\alpha_{\mu\lambda}
  \Gamma^\lambda_{\nu\beta}-\Gamma^\alpha_{\nu\lambda}
  \Gamma^\lambda_{\mu\beta},\eqno(1.5)$$
and the torsion tensor is
  $$\eqalign{T_{\rho\sigma}{}^\mu&=\Gamma^\mu_{\rho\sigma}-
  \Gamma^\mu_{\sigma\rho}\cr &=2\Gamma^\mu_{[\rho\sigma]}.}
  \eqno(1.6)$$
Thus, the curvature and torsion have a similar status as tensors
which characterize a specified connection.  Special relativity
posits a spacetime connection for which both tensors vanish;
the transition from special to general relativity may be
thought of as allowing for the dynamics of a nonzero curvature,
while constraining the torsion to vanish. {}From a point of view which
takes the connection as an independent variable, this restriction
seems somewhat arbitrary (although it is nevertheless possible, by
judicious choice of Lagrangian, to make the torsion nonpropagating
or even vanishing).
We are therefore led to consider theories in which both the
curvature and torsion are determined dynamically by the response
of the metric and connection to matter fields.

The introduction of additional propagating degrees of freedom
opens the possibility that such a theory could
lead to observable deviations from general relativity.
Experiments in the solar system and in binary pulsar 1913+16 offer
strong evidence that the metric must not deviate too far from the
form specified by Einstein's equations [7].  The  situation with
respect to torsion is less clear, as the literature contains
various different proposals for what the dynamics of torsion
could be.

Our goal in this paper is to determine whether there are any
observational consequences of propagating torsion which are
relatively independent of any specific gravitational model.
To that end, we discuss
possible actions for torsion and its interaction with matter fields
such as those in the standard model of particle physics.  In these
theories we construct a free Lagrangian from powers and derivatives
of the torsion, and couple ``minimally'' to matter through the
covariant derivative.  We find that
there is only a small range of models possible without placing
arbitrary restrictions on the dynamics.  In these models only a
single mode interacts with matter, either a massive scalar
or a massive spin-1 field, and each model is parameterized by two
constants with the dimensions of mass.  In this paper we concentrate
on the scalar theory, which is related to several different proposals
found in the literature.  We discuss what regions of parameter space
are excluded by laboratory and astrophysical data.  A reasonable
expectation, however, would be for each of the two mass parameters to
be of order the Planck scale; such a choice is a safe distance away
from the regions excluded by experiment.
We conclude that, while there are reasons to expect that the
torsion degrees of freedom exist as propagating fields,
there is no reason to expect any observable signature
from torsion.

\noindent{\bf II. Lagrangians}

Since our goal is to search for potentially observable
consequences of a theory of gravity with propagating torsion,
and not to construct the full theory itself, we shall limit our
attention to the dynamics of torsion in a background
spacetime with the Minkowski metric, $g_{\mu\nu}=\eta_{\mu\nu}$.
(Actions for gravity with
propagating torsion, but without couplings to matter, have been
studied in [8,9].)  Furthermore, we shall not worry
about the renormalizability of the torsion sector, since quantization
of the full gravity theory is beyond our reach anyway.  Nevertheless,
we shall keep in mind known quantum effects (such as anomalies)
in the matter sector.  Thus, we are interested in theories
defined by Lagrangians of the form
  $${\cal L}={\cal L}_T+{\cal L}_I+{\cal L}_M,\eqno(2.1)$$
where ${\cal L}_T$ is the part of the Lagrangian containing only
torsion fields, ${\cal L}_M$ is the matter Lagrangian, and
${\cal L}_I$ defines interactions between torsion and matter.

We turn first to the construction of ${\cal L}_I$, which
involves an unavoidable ambiguity.
In a spacetime with a metric, we may always decompose a
metric-compatible connection (which is all we shall consider) into a
torsion-free Christoffel piece plus a torsion-dependent piece.  In
terms of components we write
  $$\Gamma^\alpha_{\mu\nu}=\left\{\matrix{\alpha\cr
  \mu\nu\cr}\right\}+{1\over 2}\left(T_{\mu\nu}{}^\alpha
  -T_\nu{}^\alpha{}_\mu
  +T^\alpha{}_{\mu\nu}\right)\eqno(2.2)$$
where the Christoffel symbols $\left\{\matrix{\alpha\cr
\mu\nu\cr}\right\}$ are given by the familiar formula
  $$\left\{\matrix{\alpha\cr \mu\nu\cr}\right\}
  \equiv{1\over 2}g^{\alpha\beta}\left(\partial_\mu g_{\nu\beta}
  +\partial_\nu g_{\beta\mu}-\partial_\beta g_{\mu\nu}\right).
  \eqno(2.3)$$
The transformation properties of (2.3) are those of a connection,
just as in conventional general relativity.  Therefore the
Christoffel covariant derivative
  $$\widetilde{\nabla}_\mu X^\alpha\equiv \partial_\mu X^\alpha
  +\left\{\matrix{\alpha\cr \mu\lambda\cr}\right\}X^\lambda
  \eqno(2.4)$$
is a well-defined tensor.  Hence, when we write down the equations
describing a theory, each appearance of the covariant
derivative $\nabla A$ of a tensor field $A$ is identical to
the Christoffel
derivative $\widetilde{\nabla}A$ plus interactions between the torsion
and $A$.  A theory with torsion is thus equivalent to a theory without
torsion, plus an extra tensor field with certain couplings.  (The fact
that the torsion contribution to the connection is a tensor means that
it is impossible, when the torsion is nonzero, to choose coordinates
in which the coefficients $\Gamma^\alpha_{\mu\nu}$ vanish.)
In other words, since the torsion transforms as an ordinary
tensor field, its status as a ``gravitational'' field
characterizing the geometry of spacetime has the nature of a semantic
distinction which we may or may not choose to make.  Nevertheless, we
will suppose here that a theory of gravitation in which the
connection is an independent variable will
predict the existence of such a field (whatever we choose to call it),
and that there are certain natural interactions for the field to have,
given its origin as part of the covariant derivative on the tangent
bundle.

With this philosophy in mind, the interaction between torsion and
matter fields is straightforward to derive [5,10].  For scalar fields,
the covariant derivative is equal to the partial derivative and
hence does not involve the connection; there is therefore no
interaction between scalars and torsion.  The same result holds
for gauge fields, although the reasoning is somewhat more subtle.
(We are now speaking classically; quantum effects will change
the situation, as we discuss below.)
We consider for simplicity an abelian gauge field $A_\mu$.
The only gauge invariant derivative of $A_\mu$ we may take is the
field strength tensor $F_{\mu\nu}$, which is defined as an
exterior derivative: $F\equiv({\rm d}A)$.  In
components this is the antisymmetric {\it partial} derivative,
  $$F_{\mu\nu}\equiv ({\rm d}A)_{\mu\nu}
  =\partial_\mu A_\nu -\partial_\nu A_\mu\ .\eqno(2.5)$$
While this clearly does not involve the torsion, the mistake is
sometimes made of defining the field strength as the antisymmetric
{\it covariant} derivative.  Such a definition suffices when the
torsion vanishes, as (2.5) is then recovered; however, with
nonzero torsion the antisymmetric covariant derivative induces
a gauge non-invariant interaction between the torsion and $A_\mu$.
There is in fact no reason for such a term to exist, as (2.5) is
the correct definition even in curved space (for a full
discussion, see [10]).

There is, in contrast, a direct interaction between torsion and
fermions.  The covariant derivative of a spinor field requires the
introduction of the tetrad formalism and a spin connection, which
we have been avoiding for simplicity; however, we can
transcribe the result, which can be found in the literature
[5].  In the presence of torsion the free Dirac Lagrangian
for a massive fermion $\psi$ can be decomposed into a torsion-free
part ${\cal L}_{TF}$ plus an interaction term ${\cal L}_I$:
  $$\eqalign{{\cal L}_{Dirac}&={i\over 2}\left(
  \nabla_\mu\bar\psi\gamma^\mu\psi
  -\bar\psi \gamma^\mu\nabla_\mu\psi\right)-m\bar\psi \psi \cr
  &={\cal L}_{TF}+{i\over 8}T_{\mu\nu\lambda}
  \bar\psi\gamma^{[\mu}\gamma^\nu\gamma^{\lambda]}\psi \ ,\cr}
  \eqno(2.6)$$
where ${\cal L}_{TF}=i\bar\psi\gamma^\mu\partial_\mu\psi
-m\bar\psi\psi$ and we have set $g_{\mu\nu}=\eta_{\mu\nu}$.
Since the gamma matrices are antisymmetrized, only the completely
antisymmetric part of the torsion tensor enters the interaction.
It is useful to define the vector which is dual to this
antisymmetric part:
  $$T^\sigma\equiv{1\over{3!}}\epsilon^{\mu\nu\lambda\sigma}
  T_{\mu\nu\lambda} \ , \eqno(2.7)$$
which can be inverted to yield $T_{[\mu\nu\lambda]}=
\epsilon_{\sigma\mu\nu\lambda}T^\sigma$.  We can then use
the identity $\gamma^\mu\gamma^\nu\gamma^\lambda
\epsilon_{\mu\nu\lambda\sigma}=(i3!)\gamma_\sigma\gamma_5$ to
write the interaction as
  $${\cal L}_I ={3\over 4}T_\mu j^\mu_5 \ ,\eqno(2.8)$$
where we have used the conventional definition of the fermion
axial vector current, $j^\mu_5\equiv \bar\psi\gamma^\mu\gamma_5\psi$.

We see that the entire interaction between torsion and matter
reduces to a coupling of the axial vector current to a torsion
(pseudo-)vector
$T_\mu$.  In constructing the torsion-only Lagrangian ${\cal L}_T$,
we shall therefore confine our attention to this single vector.
While the other components of $T_{\mu\nu}{}^\lambda$ may interact
with this vector, they do not couple directly to matter, and are
therefore unlikely to yield observable effects.  (In what follows
we shall treat the vector $T_\mu$ as a fundamental field with
respect to which we vary the action to obtain equations of
motion.  In the full theory the fundamental fields would be
a vierbein and spin connection.)

It is straightforward to write down a Lagrangian for the torsion
vector which contains all possible terms of no higher than
second order in $T_\mu$ or derivatives of $T_\mu$.  We may
express it as
  $${\cal L}_T=a\partial_{[\mu}T_{\nu]}\partial^{[\mu}T^{\nu]}
  +b(\partial_\mu T^\mu)^2 +cT_\mu T^\mu\ .\eqno(2.9)$$
A term involving the symmetric part of $\partial_\mu T_\nu$ can
be absorbed, after integration by parts, into the first two terms
above.\footnote{$^3$}{We have written (2.9) in terms of
$\partial_\mu$ rather than $\nabla_\mu$, which are not equal
when the torsion is nonvanishing.  This is justified since the
covariant derivative $\nabla_\mu T_\nu$ is simply the partial
derivative $\partial_\mu T_\nu$ plus interactions between
$T_\mu$ and other components of the torsion tensor, which we
are neglecting by hypothesis.}
It is not possible to eliminate $a$, $b$ or $c$ by a
field redefinition, since the interaction (2.8) contains no
arbitrary constants.
We recall that a vector field describes four degrees of
freedom, which can be thought of as a single spin-0 field plus
the three polarization modes of a spin-1 field.  A simple
calculation reveals that it is impossible
for both the scalar and the spin-1 components to simultaneously
exist as propagating degrees of freedom once we demand that the
Hamiltonian of the theory be bounded below.
In our notation, this
means that the sign of $c$ must be negative for the
scalar to be non-tachyonic, but positive for the spin-1 field
to be non-tachyonic.  Hence, in order for our theory to be
well-defined, we may choose the parameters $a$, $b$ and $c$ such
that either the scalar or the spin-1 modes propagate, but not
both.\footnote{$^4$}{We are, of course, using classical language;
in a quantum theory we would say that it is impossible to have
four propagating degrees of freedom without involving unphysical
ghosts.  The problem is present, however, even at the classical
level.}  Setting $a=0$, $b\ne 0$ results in a
theory with a propagating scalar, while setting $b=0$, $a\ne 0$
corresponds to a massive spin-1 field.  We shall look at the
spin-0 theory more closely in the next section.  An analysis
similar to that given below should be able to provide
constraints on the corresponding spin-1 theory.

The theories we consider differ from the conventional
Einstein-Cartan theory, defined by
  $${\cal L}_{EC}=-M_P^2R\ ,\eqno(2.10)$$
where $R$ is the Ricci scalar $R=g^{\mu\nu}
R^\sigma{}_{\mu\sigma\nu}$ and $M_P$ is the Planck mass.
Upon decomposing ${\cal L}_{EC}$ in terms of the metric and
torsion, we find that $T_\mu$ enters only algebraically:
  $${\cal L}_{EC}=-M_P^2 T_\mu T^\mu +
  {\rm non-}T_\mu{\rm\ terms}. \eqno(2.11)$$
Einstein-Cartan theory, therefore, corresponds to the choices
$a=0$, $b=0$, and $c=-M_P^2$ in our equation (2.9).
Since the interaction term (2.8) is also free of derivatives
of $T_\mu$, varying the Einstein-Cartan
action with respect to $T_\mu$ yields
the constraint $T^\mu=(3/8M_P^2)j^\mu_5$.  This constraint can
then be substituted back into the Lagrangian, resulting in a new
four-fermion interaction (suppressed by two powers of the Planck
mass).  It should be clear that this choice of action leads to
no interesting long-range forces; however, we believe that there
is no good reason to limit our attention to the Einstein-Cartan
action.  If the connection is our fundamental variable, it is
unnatural to restrict the torsion degrees of freedom such that
they do not propagate, unless it is found that theories with
torsion are internally inconsistent or in conflict with experiment.
We therefore turn to exploration of such a theory.

\noindent{\bf III. Consequences}

The spin-0 Lagrangian ${\cal L}_0$ results from setting $a=0$
in (2.9), and adding the interaction given by (2.8):
  $${\cal L}_0=b(\partial_\mu T^\mu)^2 +cT_\mu T^\mu
  +{3\over 4}T_\mu j^\mu_5 \ .\eqno(3.1)$$
To make the scalar nature of this theory more explicit, we
can consider the equivalent expression ${\cal L}^\prime =
{\cal L}_0+{\cal L}_\lambda$, where
  $${\cal L}_\lambda = -b\left(\lambda-\partial_\mu T^\mu
  \right)^2\ .\eqno(3.2)$$
Here, $\lambda$ is a field which functions as a Lagrange
multiplier.  Varying with respect to $\lambda$ yields the
constraint $\lambda=\partial_\mu T^\mu$; substituting back
into (3.2), we find that ${\cal L}_\lambda$ vanishes,
so that ${\cal L}_0$ and ${\cal L}^\prime$ define identical
theories.  However, we may choose instead to keep $\lambda$
in the Lagrangian, and after an integration by parts we
obtain
  $${\cal L}^\prime=cT_\mu T^\mu +{3\over 4}T_\mu j^\mu_5
  -b\lambda^2 -2bT^\mu\partial_\mu\lambda\ .\eqno(3.3)$$
In this version there are no derivatives of the torsion,
and it is $T_\mu$ which functions as a Lagrange
multiplier (although the physics is, of course, still the same
as (3.1)).  Variation with respect to $T_\mu$ yields the
constraint
  $$T_\mu = {b\over c}\partial_\mu\lambda -{3\over{8c}}
  j_{\mu 5}\ .\eqno(3.4)$$
We can insert this back into (3.3) to obtain an expression
solely in terms of $\lambda$ and $j^\mu_5$.  To make things
look conventional, we define $m^2\equiv -c/b$,
$f\equiv(-8c/9)^{1/2}$, and $\phi\equiv(-2b^2/c)^{1/2}\lambda$,
which gives
  $${\cal L}^\prime={1\over 2}(\partial_\mu\phi)^2 -{{m^2}
  \over 2}\phi^2 + {1\over{2f}}\phi\partial_\mu
  j^\mu_5 +{1\over{8f^2}}j_{\mu 5}j^\mu_5\ .\eqno(3.5)$$
Thus, this choice of torsion action is equivalent to a
conventional pseudoscalar field with mass $m$
coupled to the divergence of the
axial vector current, along with an induced four-fermion
interaction.  Notice that we must require that $b>0$ and
$c<0$ to guarantee that the (mass)$^2$ of $\phi$ be positive
and that $f$ be real.

While the Lagrangian (3.5) specifies the entire classical
theory, quantum effects (in the matter sector) will lead to
additional interactions.  Specifically, there will be an
interaction with gauge bosons, mediated by triangle diagrams,
due to the chiral anomaly
in the current $j^\mu_5$.  The torsion scalar couples to the
divergence of $j^\mu_5$, given by
  $$\partial_\mu j^\mu_5 = {{N_f \alpha}\over {4\pi}}F_{\mu\nu}
  \widetilde F^{\mu\nu} + \sum_i m_{\psi_i}\bar\psi_i \gamma_5
  \psi_i\ ,\eqno(3.6)$$
where $F_{\mu\nu}$ is the gauge field strength (we limit our
attention to electromagnetism), $\widetilde F^{\mu\nu}\equiv
{1\over 2}\epsilon^{\mu\nu\rho\sigma}F_{\rho\sigma}$ is
its dual, $N_f$ is the sum of the electric charges  of
the fermions $\psi_i$, $\alpha$ is the fine
structure constant, and $m_{\psi_i}$ is the
mass of $\psi_i$.  Equation (3.6) induces
an effective interaction Lagrangian ${\cal L}_{\phi
F\widetilde F}$ between torsion and gauge fields:
  $${\cal L}_{\phi F\widetilde F}={{N_f\alpha}\over{8\pi f}}
  \phi F_{\mu\nu}\widetilde F^{\mu\nu}\ .\eqno(3.7)$$
Hence, while there is no interaction between torsion and gauge
fields at the classical level, quantum effects (in the form of
the chiral anomaly) induce a coupling, which may help to
place constraints on the theory.

We now discuss potentially observable consequences of
the theory defined by (3.5).  In doing this, we may treat the two
scales $m$ and $f$ as completely free parameters, and ask what
values lead to detectable effects.  However, there is good reason
to expect that both $m$ and $f$ should be of order the Planck mass
$M_P$.  Indeed, if the Lagrangian includes the Ricci scalar
constructed from the connection (the usual Einstein-Cartan choice)
as well as pure torsion terms, there is
automatically a contribution to $m^2$ of order $M_P^2$, as evidenced
by (2.11).  Thus, it would require a certain degree of fine tuning
for $m$ to be much less than $M_P$.  Similarly, since $f$ can be
though of as $m$ times a dimensionless constant which we would
expect to be of order unity, $m\sim f\sim M_P$ is a reasonable
expectation.  However, the resulting theory clearly leads to no
observable phenomena.  A scalar particle with a mass
$M_P$ would not be produced in any conceivable experiment;
furthermore, it would decay into fermions or gauge bosons with
a lifetime $\tau\sim f^2/m^3 \sim M_P^{-1}$ [11], so any
bosons produced in the early universe would quickly decay away.
Thought of as a classical field, the effective range of $\phi$
is also given by $l\sim M_P^{-1}\sim 10^{-35}{\rm ~cm}$; no realistic
source would give rise to a long-range field which might be observed.
Hence, it is no surprise that torsion has not been detected by
any experiment, and the lack of such detection should not be taken
as strong evidence that torsion plays no role in the fundamental
theory.

While it is reasonable to locate the parameters of the torsion
theory in the Planck regime, it is nevertheless possible that
they lie at much lower energies.  Fortunately, the resemblance of
the interactions in (3.5) to previously studied theories allows
us to readily catalogue the limits on $m$ and $f$.  These are
presented in Fig.~1.  It is clear that the constraints, while
interesting, do not approach the Planck scale.

The four-fermion
interaction $-(1/8f^2)j_{\mu 5} j^\mu_5$ allows us to place
a limit on $f$, independent of $m$.  Such terms have been
studied in the context of composite models for quarks and leptons,
in which the effective theories often include four-fermi
interactions induced at the compositeness scale [12].  The interaction
between two axial vector currents is parameterized by a mass scale
$\Lambda_{AA}$, related to our $f$ by $f^2=\Lambda_{AA}^2/32\pi$.
Constraints on $\Lambda_{AA}$ arise from electron-positron
annihilation experiments, in which the four-fermion interaction
contributes to a charge asymmetry over and above that expected in the
standard model.  The best current limits on such an
interaction come from $e^+e^-\rightarrow q\bar q$ observations at
TRISTAN [13].  These
experiments yield a limit $\Lambda_{AA}\geq 3{\rm ~TeV}$, or
  $$f\geq 3\times 10^2{\rm ~GeV}\ .\eqno(3.8)$$
This limit is possible to circumvent, however, by adding a fundamental
four-fermion interaction to the initial Lagrangian (3.1).  The
effect of such a term would be to alter the relation between
the torsion coupling constant $f$ and the parameter $\Lambda_{AA}$
governing the strength of the four-fermion interaction, without
changing any of the dynamics of the torsion scalar $\phi$.  With
this in mind, we have indicated the limit (3.8) by a dashed line
in Fig.~1.

The interaction of $\phi$ with fermions also leads to constraints
from laboratory experiments.  The most effective limits come from
searches for neutral bosons in $\Upsilon\rightarrow \phi +\gamma$
and $J/\psi \rightarrow \phi+\gamma$ events [14].  Current
data enable us to place the limit
  $$m\geq 1\times 10^{-3}{\rm \ GeV}\ ,\eqno(3.9)$$
which however is only valid for $f\leq 10^3$~GeV.  This enables us
to rule out the bottom left corner of Fig.~1.

Astrophysical effects of the $\phi$ bosons lead to constraints in
the $m - f$ plane for somewhat higher values of $f$.
The interactions of $\phi$, as
specified by (3.5) and (3.7), are precisely those of a pseudo-Goldstone
boson (PGB) resulting from the spontaneous breakdown of a global
symmetry at a scale $f$, followed by the explicit breakdown
of the symmetry at a scale $\Lambda=\sqrt{mf}$.  We can use
this similarity to our advantage, by applying astrophysical
constraints on PGB's to the theory at hand;
however, we wish to emphasize that the resemblance is (as far
as we know) purely coincidental.  There is no spontaneously
broken symmetry for which $\phi$ is the pseudo-Goldstone boson;
indeed, some of the essential physics is different.

Frieman and Jaffe [11] have presented a comprehensive list of
astrophysical constraints on PGB's; we summarize the processes
most relevant to our purposes.  In the regime left unconstrained by
ground-based experiments, the most effective bounds come from
processes in which PGB's lead to energy loss in stars.  Briefly,
there may be a range of parameter space in which the mass of
$\phi$ is low enough that it can be produced in a stellar
interior and the coupling to ordinary matter is sufficiently
strong that the rate of production is significant, while at
the same time sufficiently weak that the PGB will often escape
without further interaction, providing an additional channel
for energy loss from the stellar core.
Three distinct manifestations of this effect lead to
interesting bounds:  shortening the lifetime of helium-burning
(horizontal branch) stars, preventing helium ignition
in low-mass red giants,
and shortening the duration of the neutrino pulse from
supernova 1987A.  The most effective bounds come from SN~1987A; in
this case, the coupling of $\phi$ to nucleons can cool the supernova
core and noticeably decrease the duration of the neutrino burst.
This leads to the constraint
  $$m\ge 6\times 10^{-2}{\rm ~GeV}\ ,\eqno(3.10)$$
valid for $5\times 10^6{\rm ~GeV}\le f \le 10^{10}{\rm ~GeV}$.
Larger masses would not affect the supernova, since in that case
$m$ would be higher than the characteristic energy of the supernova
core.  (We are being
somewhat loose in quoting these bounds; more precise information
can be found in [11].)   The resulting gap between $f=10^3
{\rm ~GeV}$ and $f=5\times 10^6{\rm ~GeV}$ can be closed using the
effects on horizontal branch stars and red giants; the former arises
because energy loss via the Primakoff process ($\phi\rightarrow
\gamma$ by scattering off electrons or nuclei) decreases the time a
star will spend in the helium-burning stage, while the latter results
when bremsstrahlung ($\phi$ production in electron-nucleus scattering)
allows red giant cores to cool sufficiently to prevent helium
ignition.  Taken together, these phenomena lead to the bound
  $$m\ge 2\times 10^{-5}{\rm ~GeV}\ ,\eqno(3.11)$$
applicable for all $f\leq 10^9$~GeV.  The astrophysical and
laboratory constraints are summarized in Fig.~1.
We also note that a seperate set of astrophysical effects are
expected for very low-mass or long-range fields; details may be
found in [15].

It is important to note that some of the most restrictive
cosmological limits on PGB's have no analogue in the torsion
theory --- specifically, those from production of PGB's in the
early universe via cosmic string decay and vacuum misalignment [11,16].
In each of these cases the constraint arises because, in certain
regions of parameter space, PGB's dominate the energy density of
the Universe: $\Omega_\phi h^2 \geq 1$, where
$\Omega$ is the density parameter of a Robertson-Walker universe
and $h$ is the Hubble constant in units of 100~km/sec/Mpc.
Determinations of the age of the universe imply that $\Omega_{tot}
h^2 \leq 1$, leading to limits on $m$ and $f$.  However, the
production of particles by string decay or vacuum misalignment
depends intimately on the nature of $\phi$ as a pseudo-Goldstone
boson, the angular degree of freedom in a tilted Mexican hat potential
resulting from spontanteous symmetry breaking.  The torsion theory,
in contrast, leads to neither strings nor vacuum misalignment; hence,
these constraints are inapplicable.  (Conventional thermal production
of $\phi$ particles can contribute significantly to the density
parameter, but only in a small region of
parameter space which is already excluded by the argument from
SN1987A.)

\noindent{\bf IV. Discussion}

We have discussed the empirical constraints on a theory of gravity
with a propagating scalar torsion degree of freedom.  This theory
arises naturally out of a simple set of assumptions, and the
associated spin-0 particle is likely to appear in a wide variety
of Lagrangians with propagating torsion.  Whereas a natural
expectation would be for the mass scales characterizing the theory
to approach the Planck scale, the region of parameter space
accessible to experiment is naturally at much lower energies.
{}From this point of view, it is not surprising that torsion-free
general relativity is successful at explaining known observational
data.

It is interesting to contrast the theory examined here with other
proposals in the literature.  Sezgin and van~Nieuwenhuizen [9] have
studied tacyhon-free gravitational theories, and present five
Lagrangians involving the metric and torsion.  Their theory
1 is Einstein-Cartan theory; theories 2, 4 and 5 propagate the
massive scalar particle described in this paper (as well as
others); and theory 3 propagates the massive vector particle
correponding to the choices $b=0$, $a\neq 0$ in our Eq.~(2.9).
They did not discuss observational constraints on their theories,
or couplings to matter.

Meanwhile, several papers have considered theories involving a
scalar torsion field coupled to $\partial_\mu j^\mu_5$ or
$F_{\mu\nu}\widetilde F^{\mu\nu}$.  An early version of such a
theory was proposed by Novello [17], who attempted to couple
torsion to electromagnetism in a gauge-invariant fashion.  He
argued that this was possible if the dual torsion vector was
restricted to be the gradient of a scalar,
$T_\mu=\partial_\mu\phi$.  A similar proposal was examined in
greater detail by De~Sabbata and Gasperini [18].  They computed
the photon propagator in QED with a {\it constant}
torsion background, and found that the result was equivalent to
the introduction of an effective interaction $T_\mu A_\nu
\widetilde F^{\mu\nu}$.  On the basis of this result and the
desire to preserve gauge invariance, they imposed the
restriction that the torsion vector be the gradient of a scalar.
The component (2.11) of the Einstein-Cartan Lagrangian involving
$T_\mu$ then becomes a conventional kinetic term for a scalar
field, $\partial_\mu\phi \partial^\mu\phi$.  Since the scalar
field only appears in the form $\partial_\mu\phi$, it is massless,
and the torsion can lead to long-range interactions.  In a similar
vein, Hammond [19] has proposed an antisymmetric two-index torsion
potential $\psi_{\mu\nu}$, related to the torsion tensor by
$T_{\mu\nu\sigma}=\partial_{[\sigma}\psi_{\mu\nu]}$ and coupled
to electromagnetism through an interaction of the form
$F^{\mu\nu}\psi_{\mu\nu}$.
Finally, Duncan, Kaloper and Olive [20] have examined
Einstein-Cartan theory with the addition of a constraining term
$\phi\partial_\mu T^\mu$, where $\phi$ is treated as a Lagrange
multiplier.  This technique, without imposing any external
restrictions on the form of $T_\mu$ or $\phi$, introduces a
propagating massless scalar field which couples to $\partial_\mu
j^\mu_5$.

There are therefore two important distinctions between these
investigations of the consequences of scalar torsion theories
and the approach advocated in this paper.  First,
in defining the Lagrangian, we have been able to describe a
propagating scalar torsion degree of freedom without imposing
external restrictions on the form of the torsion tensor, nor
changing the degree-of-freedom content by
introducing auxilliary fields as Lagrange multipliers.  (In this
sense our approach is that of [9].)  The second
distinction, which is a direct consequence of the first, is that we
have found that the resulting scalar should be massive, and indeed
with a mass of order the Planck scale.  This is not to say that
the theories described above are necessarily incorrect; we
believe that the approach followed in this paper is simple and
natural, but at the current level of understanding this is
purely a matter of taste.

The picture of torsion as an extremely short-range field runs
somewhat counter to the intuitive conception of torsion as a part of
spacetime geometry.  More concretely, we are used to gauge theories
giving rise to massless, long-range fields, and the status of
torsion as part of the connection on the tangent bundle might
lead us to expect the same in this case.  This conflict with
intuition may be resolved by noticing that the torsion is a
tensor which is {\it linear} in the connection.  It therefore
becomes possible to
construct gauge invariant interactions which give
a mass to some of the connection degrees of freedom.  This is
in contrast with the pure metric theory, or with gauge theories
on internal vector bundles, where all gauge invariant terms
involve the curvature tensor, constructed from derivatives
of the fundamental fields.  Thus,
despite its origin as part of the geometry of spacetime, the
physical manifestation of torsion can be significantly different
from that of other ``geometrical'' fields.

The possible existence of torsion is of interest both in the
construction of quantum theories of gravity and in the experimental
search for deviations from general relativity.  The important lesson
of this paper is that the absence of effects of torsion in
experiments should not lead us to discount the
possibility of torsion playing a role in the ultimate theory of
gravity.

\noindent{\bf Acknowledgments}

It is a pleasure to thank Eric Blackman, Sidney Coleman, Edward
Farhi, Roman Jackiw, and Ted Pyne for many useful discussions.
This work was supported in part by NASA under Grants No.\ NAGW-931
and NGT-50850, by the National Science Foundation under grant
PHY/92-06867, and by the U.S. Department of Energy under
contract No.\ DE-FC02-94ER40818.

\vfill\eject

\centerline{\bf References}

\item{1.} R. M. Wald, {\it General Relativity} (University of
Chicago Press, Chicago, 1984).

\item{2.} A. Ashtekar, {\it Lectures on Nonperturbative Quantum
Gravity} (World Scientific, Singapore, 1991).

\item{3.} A. Achucarro and P.K. Townsend, {\it Phys. Lett.}
{\bf B229}, 383 (1989); E. Witten, {\it Nucl. Phys.} {\bf B311},
80 (1990).

\item{4.} G. Grignani and G. Nardelli, {\it Phys. Rev. D} {\bf 45},
2719 (1992); D. Cangemi and R. Jackiw, {\it Ann. Phys.} {\bf 225},
229 (1993); J. D. Romano, {\it Geometrodynamics Vs. Connection
Dynamics}, preprint UMDGR-93-129, gr-qc/9303032 (1993); P. Peld\'an,
{\it Actions for Gravity, with Generalizations: A Review},
G\"oteborg preprint ITP 93-13, gr-qc/9305011 (1993).

\item{5.} F. W. Hehl, P. von der Heyde, G. D. Kerlick, and J. M.
Nester, {\it Rev. Mod. Phys.} {\bf 48}, 393 (1976).

\item{6.} N. Straumann, {\it General Relativity and Relativistic
Astrophysics} (Springer-Verlag, Berlin, 1984).

\item{7.} C. M. Will, {\it Theory and Experiment in Gravitational
Physics} (Cambridge Univ. Press, Cambridge, UK, 1993).

\item{8.} D. E. Neville, {\it Phys. Rev. D} {\bf 18}, 3535 (1978).

\item{9.}E. Sezgin and P. van Nieuwenhuizen, {\it Phys. Rev. D}
{\bf 21}, 3269 (1980).

\item{10.} I.M. Benn, T. Dereli, and R. W. Tucker, {\it Phys. Lett.}
{\bf 96B}, 100 (1980).

\item{11.} J. A. Frieman and A. H. Jaffe, {\it Phys. Rev. D} {\bf 45},
2674 (1992).

\item{12.} E. J. Eichten, K. D. Lane, and M. E. Peskin, {\it Phys.
Rev. Lett.} {\bf 50}, 811 (1983).

\item{13.} K. Abe {\it et al.}, {\it Phys. Lett.} {\bf 232B}, 425
(1989); I. Adachi {\it et al.}, {\it Phys. Lett.} {\bf 255B},
613 (1991).

\item{14.} J. E. Kim, {\it Phys. Rep.} {\bf 150}, 2 (1987).

\item{15.} S.M. Carroll and G.B. Field, {\it Phys. Rev. D}
{\bf 43}, 3789 (1991); D. Harari and P. Sikivie, {\it
Phys. Lett.} {\bf 289B}, 67 (1992); S. Mohanty and P.K.
Panda, preprint hep-ph/9403205 (1994); W.D. Garretson and E.D.
Carlson, preprint (1994).

\item{16.} D. Harari and P. Sikivie, {\it Phys. Lett.}
{\bf 195B}, 361 (1987);
R. Davis and P. Shellard, {\it Nucl. Phys.} {\bf B324},
167 (1989); A. Dabholkar and J. Quashnock, {\it Nucl. Phys.}
{\bf B333}, 815 (1990).

\item{17.} M. Novello, {\it Phys. Lett.} {\bf 59A}, 105 (1976).

\item{18.} V. De Sabbata and M. Gasperini, {\it Phys. Lett} {\bf
77A}, 300 (1980); V. De~Sabbata and M. Gasperini, {\it Phys.
Lett.} {\bf 83A}, 115 (1981).

\item{19.} R. T. Hammond, {\it General Relativity and Gravitation}
{\bf 23}, 1195 (1991).

\item{20.} M. J. Duncan, N. Kaloper and K. A. Olive, {\it Nucl.
Phys.} {\bf B387}, 215 (1992).

\vfill\eject

\centerline{\bf Figure Caption}
\vskip .4cm

\noindent
Figure One.  {\it Limits on parameters characterizing the pseudoscalar
torsion theory.}  This theory is specified by two constants $m$ and
$f$, each with dimensions of mass.  We have plotted the regions
excluded by astrophysical and terrestrial data, as explained in the
text.  The solid lines represent inescapable limits, while the dashed
line may be avoided by modification of the induced four-fermion
interaction.  The star in the upper right represents
$m=f=M_P=10^{19}$~GeV, which we argued was a reasonable expectation.
Clearly, the constraints are far removed from this point.

\bye